\documentclass[3p,times]{elsarticle}

\usepackage{ecrc}

\usepackage{epstopdf}

\volume{00}

\firstpage{1}

\journalname{Nuclear Physics A}

\runauth{P. Hegde}

\jid{nupha}

\jnltitlelogo{Nuclear Physics A}

\usepackage{graphicx}
\usepackage{amsmath,amssymb}

  \newcommand{\Nt}{N_\tau}
  \newcommand{\muB}{\mu_B}
  \newcommand{\muQ}{\mu_Q}
  \newcommand{\muS}{\mu_S}
  \newcommand{\veps}{\varepsilon}
  \newcommand{\bmu}[1]{\left(\frac{\mu_B}{T}\right)^#1}
  \newcommand{\ord}{\mathcal{O}}
  \newcommand{\sNN}{s_{NN}^{1/2}}

\begin{document}

\begin{frontmatter}



\title{The QCD Equation of State to $\ord(\muB^4)$ from Lattice QCD}

\author{Prasad Hegde\footnote{Email: \texttt{phegde@mail.ccnu.edu.cn.}} (for the BNL-Bielefeld-CCNU Collaboration)}
\address{Key Laboratory of Quark and Lepton Physics (MOE) \& Institute of Particle Physics, Central China Normal University, Wuhan 430079, China.}




\begin{abstract}
We present first results from a first-principles calculation of the QCD equation of state to $\ord(\mu_B^4)$, where $\muB$ is the baryon chemical potential.  We find that second-order corrections are sufficient for a large part of the freeze-out temperature and baryon chemical potential range achieved by the RHIC beam energy scan. Nevertheless, higher-order corrections are necessary to extend the validity of the equation of state down to beam energies $s^{1/2}_{NN}\sim 20$~GeV.
\end{abstract}

\end{frontmatter}



\section{Introduction}
\label{sec:intro}
Lattice results for the Equation of State (EoS) have long been used to model the hydrodynamic evolution of the thermal matter that is created in heavy-ion collisions. Although state-of-the-art results exist for the EoS at $\muB=0$~\cite{Bazavov:2014noa,Borsanyi:2013bia}, with the advent of the Beam Energy Scan (BES) experiment at the Relativistic Heavy Ion Collider (RHIC), it has become necessary to extend these results to moderately large values of the baryon chemical potential $\muB$. Specifically, values $\muB\approx$ 400-450 MeV are expected to be reached at the lowest center-of-mass energies~\cite{Cleymans:1999st}.

Unfortunately, the well-known sign problem in lattice QCD prevents a straightforward extension of the usual techniques used to calculate the EoS at $\muB=0$ to the region $\muB>0$~\cite{deForcrand:2010ys}. While a complete solution to the sign problem is not yet known, various methods have been suggested to partially overcome it. Among these, the method of Taylor expansions is the most straightforward~\cite{Gavai:2001fr, Allton:2002zi}. The method has the advantage that the only error (apart from the usual statistical error) is the one coming from the truncation of the Taylor series. Moreover, the various coefficients of the expansion bear a straightforward interpretation as either the cumulants of the various conserved charge distributions (diagonals), or as the correlations between them (off-diagonals). Because of this, they can be used to probe deconfinement~\cite{Jeon:1999gr} and they can also be measured in experiments via the moments of different hadron multiplicity distributions~\cite{Adamczyk:2014fia}.

The basic thermodynamic quantity is the pressure $p$, which at $\muB>0$ may be written as
\begin{equation}
\frac{p}{T^4} = \sum_{i,j,k=0}^\infty%
\frac{\chi_{ijk}}{i!\,j!\,k!}%
\left(\frac{\mu_B}{T}\right)^i \left(\frac{\mu_Q}{T}\right)^j \left(\frac{\mu_S}{T}\right)^k%
\longrightarrow
\sum_{n=0}^\infty c_n\left(\frac{\muB}{T}\right)^n.
\label{eq:definition}
\end{equation}

With three flavors of quarks, one has three chemical potentials. A change of basis allows us to express these in terms of conserved charge chemical potentials: baryon number, electric charge and strangeness: $(\muB,\muQ,\muS)$. Eq.~\eqref{eq:definition} is a completely general expression. However, by specializing to the case of heavy-ion collisions and taking into account the constraints coming from the initial conditions\footnote{The initial conditions are: zero net strangeness ($n_S=0$), and a fixed electric charge-to-baryon ratio ($N_p/(N_p+N_n)=r$). Using these, $\muQ$ and $\muS$ may be determined upto any given order in $\muB$~\cite{Bazavov:2012vg}.}, we can express $\muQ$ and $\muS$ in terms of $\muB$. This makes our expansion effectively one-dimensional \emph{i.e.} in the variable $\muB$. The $c_n$ coefficients, as well as $\muQ$ and $\muS$, can all be expressed in terms of the $\chi_{ijk}$; thus it suffices to calculate all the $\chi_{ijk}$ upto a certain order. Current perturbative calculations are applicable only for temperatures $T\gtrsim 350$ MeV or so~\cite{Haque:2014rua,Laine:2006cp}. Thus the calculation of susceptibilities around and just above the crossover region is a non-perturbative problem requiring the use of lattice techniques.

\section{Results}
\label{sec:results}
\begin{figure}[!htb]
\includegraphics[width=0.30\textwidth]{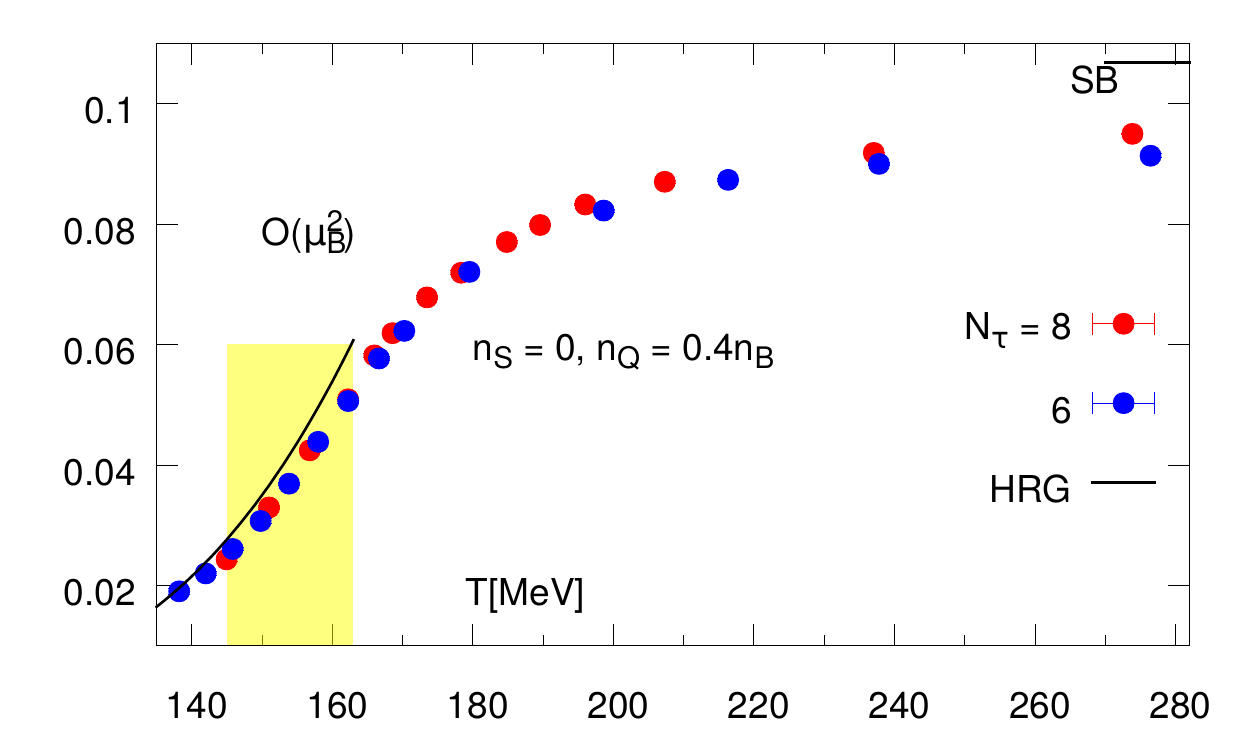}%
\hspace{0.05\textwidth}%
\includegraphics[width=0.30\textwidth]{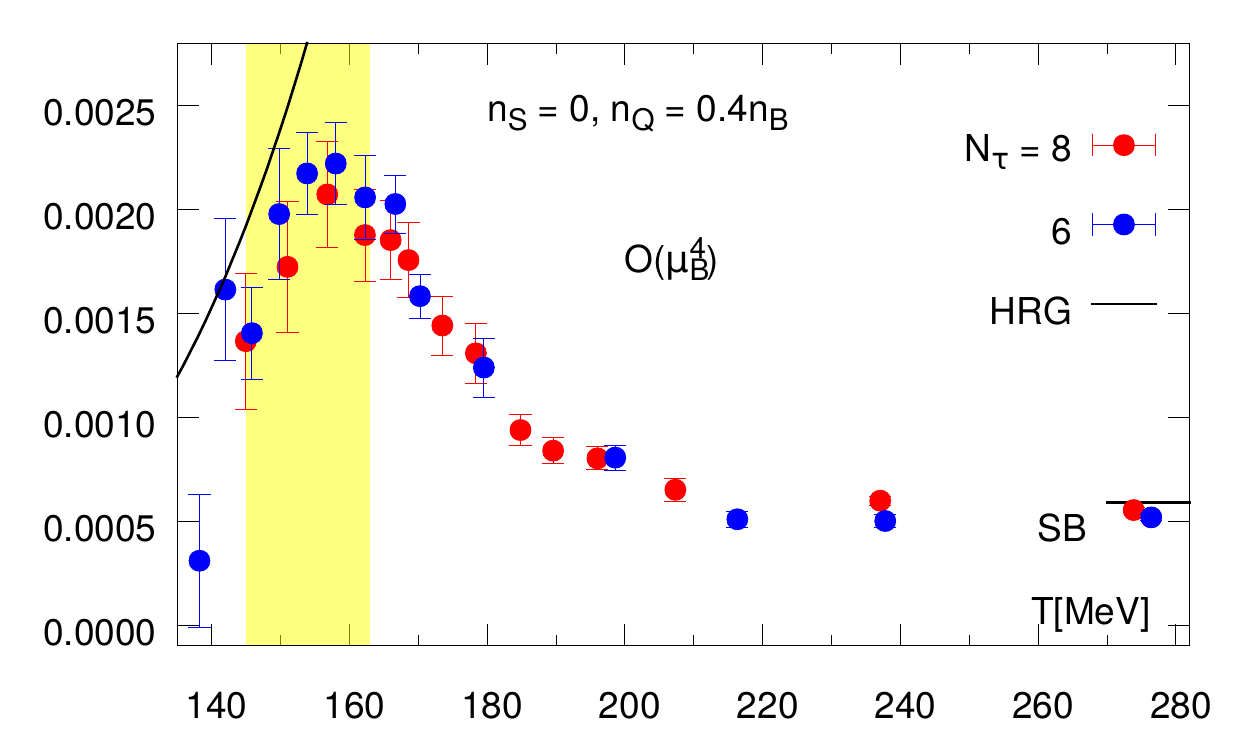}%
\hspace{0.05\textwidth}%
\includegraphics[width=0.30\textwidth]{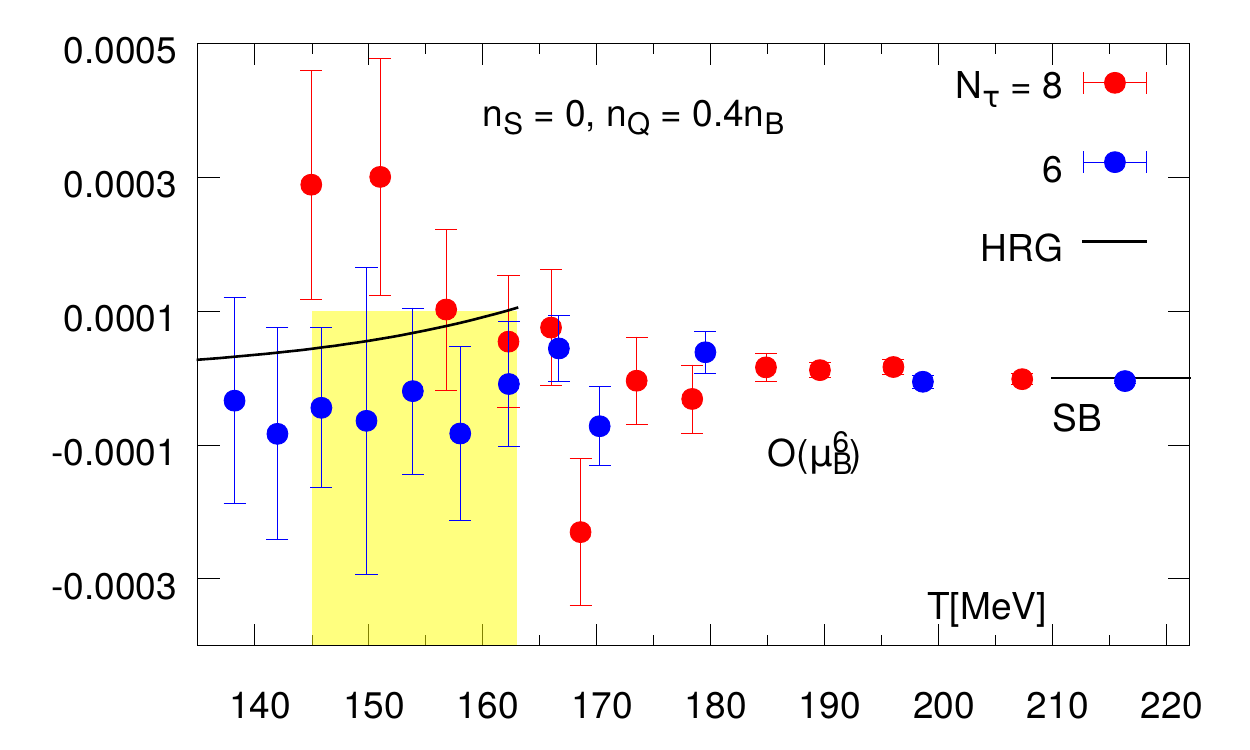}
\caption{(From left to right) Lattice results for the coefficients $c_2$, $c_4$ and $c_6$ with initial conditions appropriate to Pb-Pb collisions ($n_S=0$, $N_p=0.4(N_p+N_n)$) for $\Nt=8$ and 6. Shaded yellow bands denote the location of the chiral crossover temperature $T_c=154(9)$ MeV~\cite{Bazavov:2011nk}. Also shown are Hadron Resonance Gas (HRG) predictions below and upto the chiral crossover temperature.}
\label{fig:cn}
\end{figure}
Fig.~\ref{fig:cn} shows our preliminary results for the Taylor coefficients $c_2$, $c_4$ and $c_6$. These were calculated with staggered fermions using the state-of-the-art HISQ action~\cite{Follana:2006rc}. We computed all the susceptibilities upto sixth order at two lattice spacings, namely $a = 1/\Nt T$ with $\Nt=8$ and 6, in a temperature range 140 MeV $\lesssim T \lesssim$ 330 MeV. Since we had only two values of the lattice spacing, we did not attempt to perform a continuum extrapolation. Nevertheless we found that cutoff effects were under control, especially for $c_2$ but also for $c_4$ as well\footnote{In fact, the electric charge sector does suffer from cutoff effects at these spacings~\cite{Bazavov:2012jq}. However, since we are expanding with respect to $\muB$, the contribution of this sector is suppressed.}. Our dominant errors in fact were statistical, as Fig.~\ref{fig:cn} shows, especially for $c_6$ and to an extent for $c_4$. Even taking the large errors on $c_6$ into consideration, we still found that $|c_6| < c_4 \ll c_2$. As a result we were able to extrapolate to $\ord(\mu_B^4)$ for various observables for fairly large values of $\mu_B/T$.

\begin{figure}[!tbh]
\includegraphics[width=0.30\textwidth]{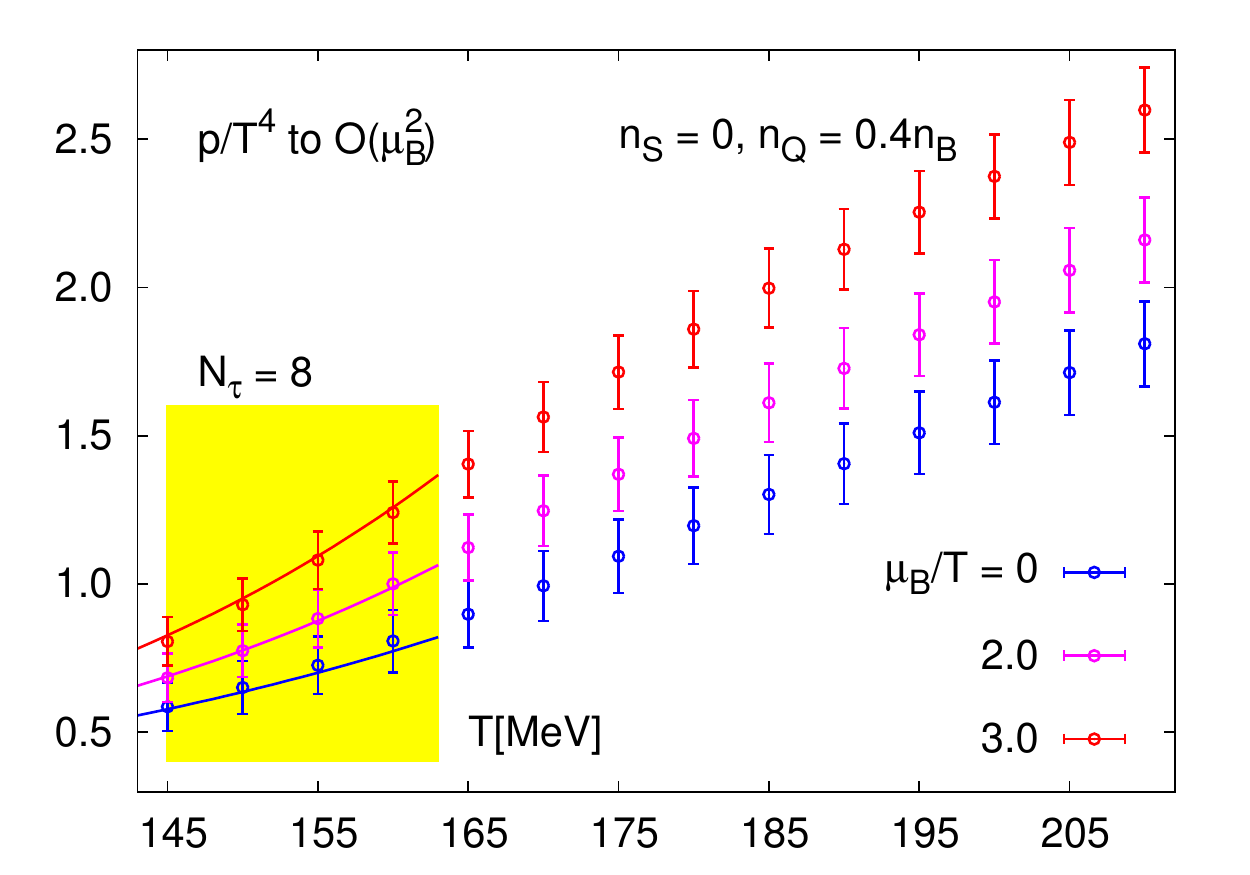}%
\hspace{0.05\textwidth}%
\includegraphics[width=0.30\textwidth]{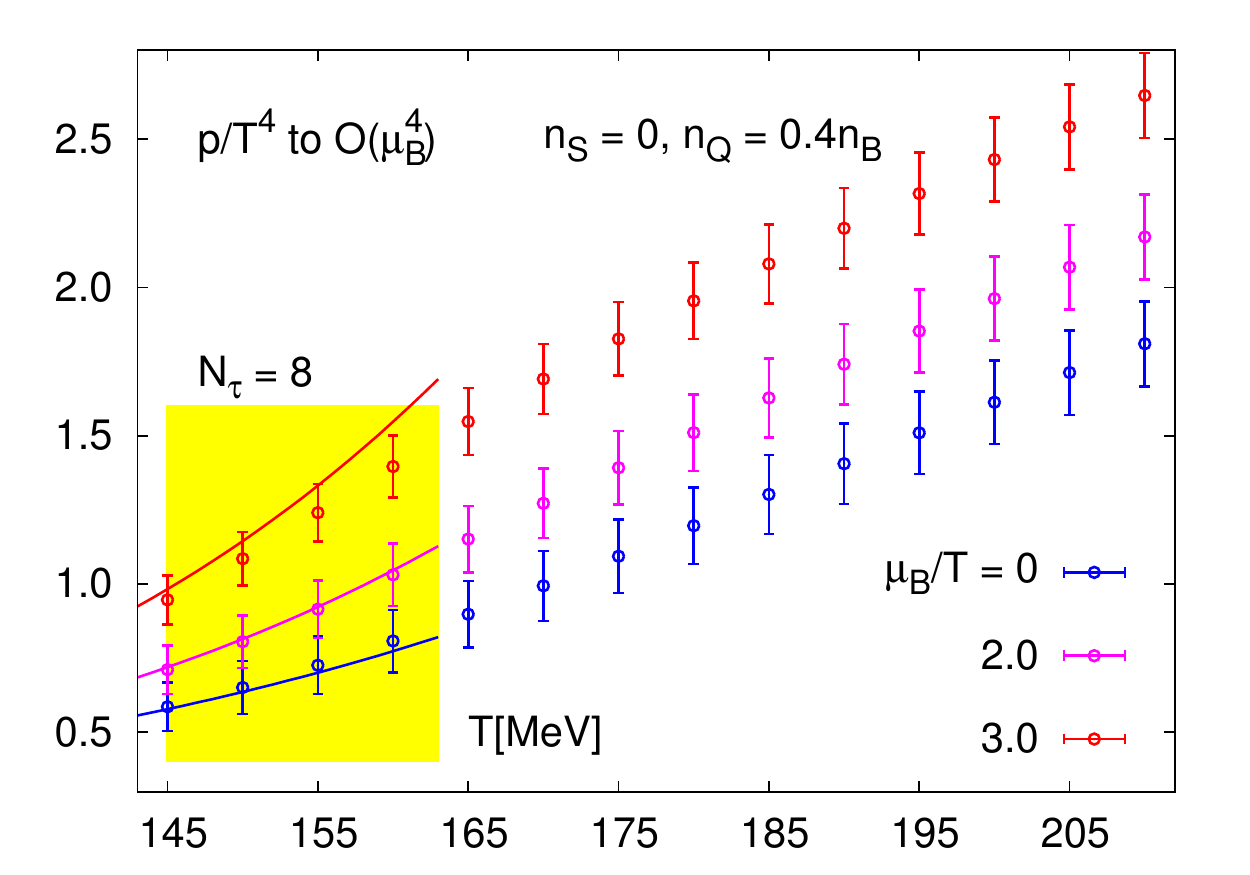}%
\hspace{0.05\textwidth}%
\includegraphics[width=0.30\textwidth]{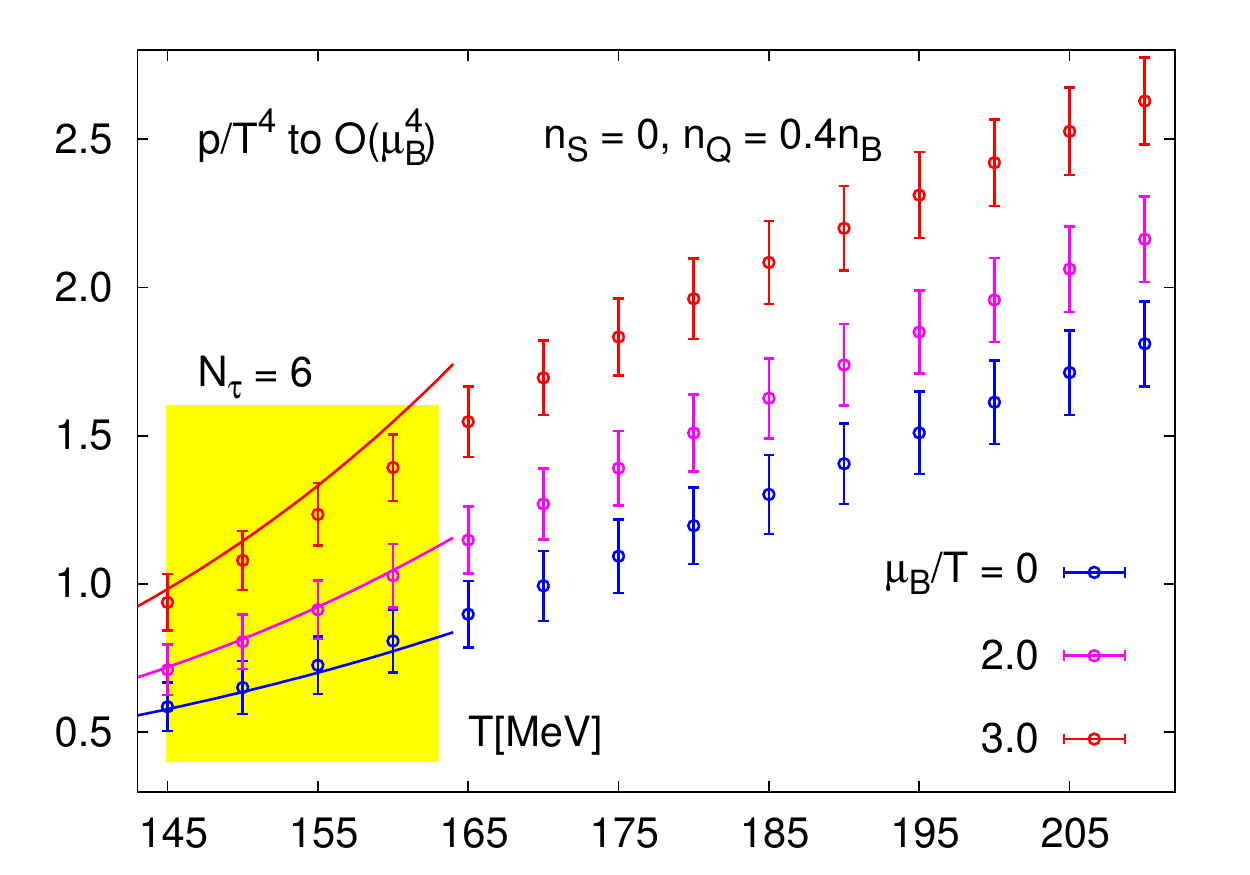}%
\caption{$p/T^4$ calculated for $\Nt=8$ upto $\ord(\mu_B^2)$ (left) and $\ord(\mu_B^4)$ (center), and to $\ord(\mu_B^4)$ for $\Nt=6$ (right). The solid colored curves for $T\leq T_c$ are the corresponding HRG results. The zeroth-order results are taken from Ref.~\cite{Bazavov:2014noa}.}
\label{fig:pressure}
\end{figure}

\begin{figure}[!t]
\includegraphics[width=0.30\textwidth]{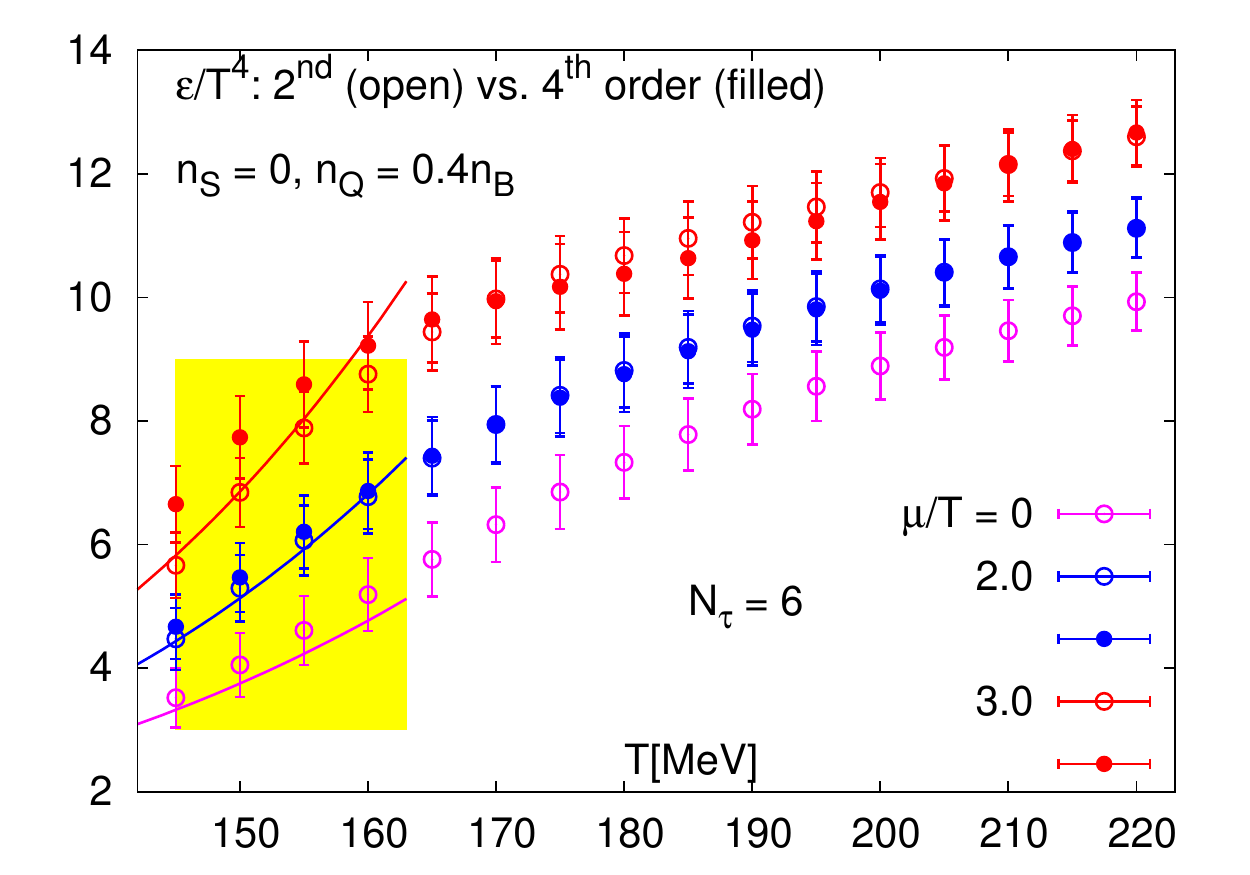}%
\hspace{0.05\textwidth}%
\includegraphics[width=0.30\textwidth]{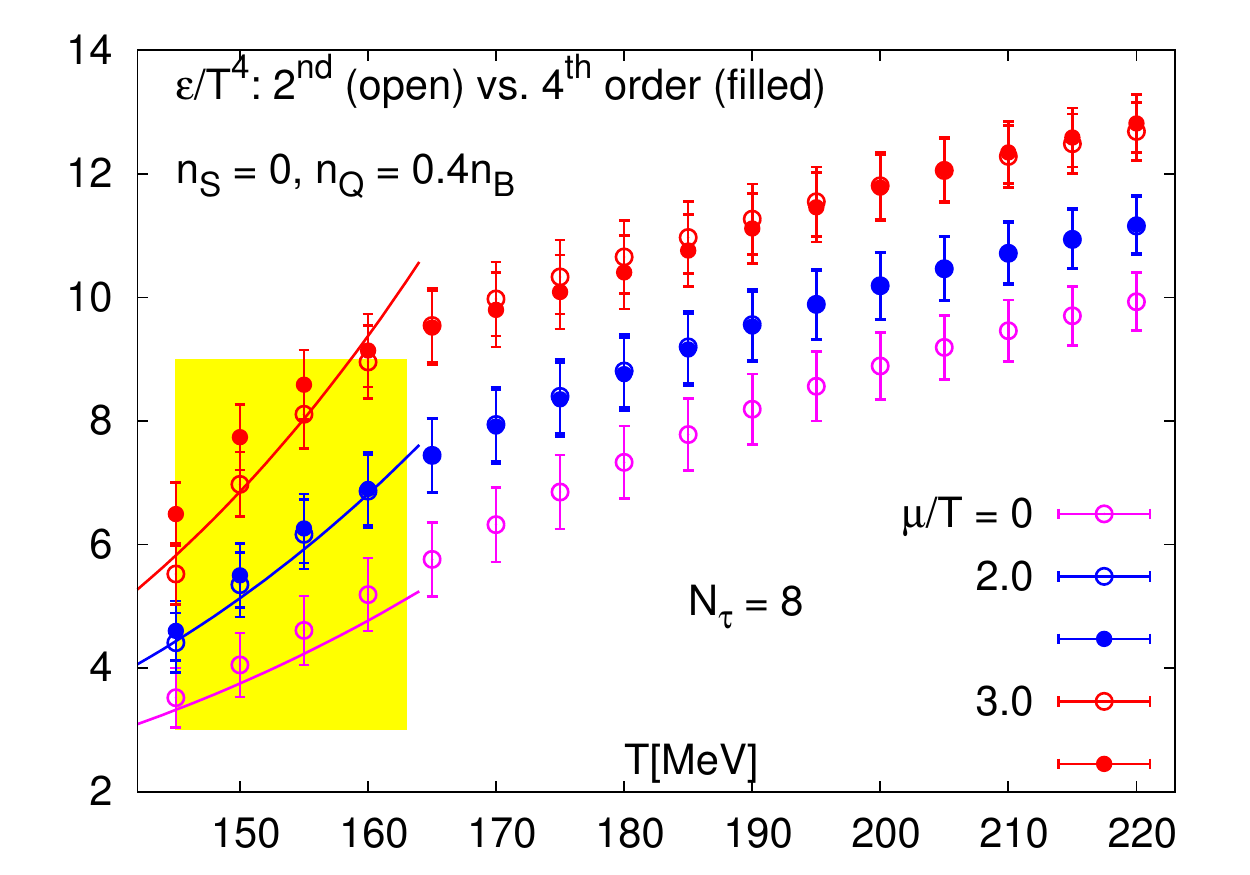}%
\hspace{0.05\textwidth}%
\includegraphics[width=0.30\textwidth]{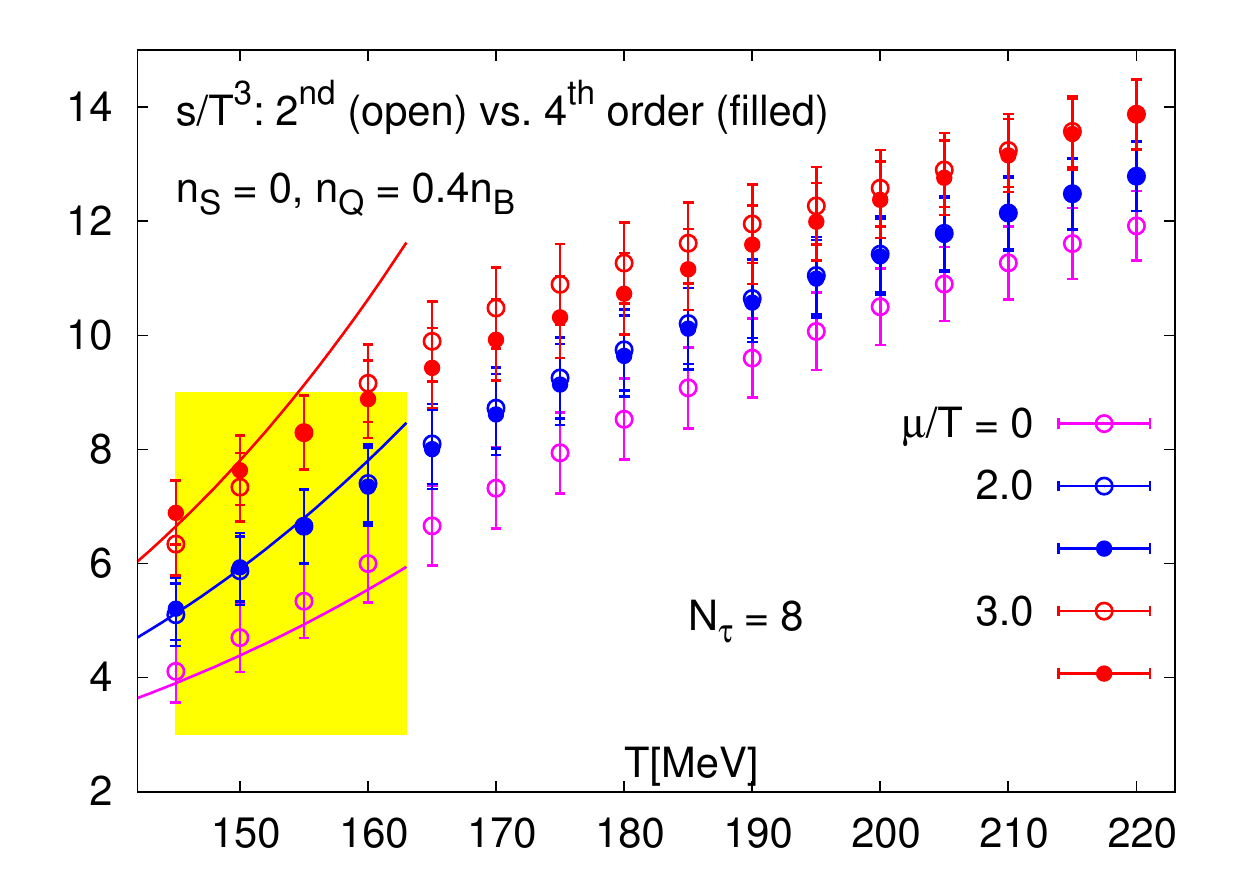}%
\caption{Comparision between second and fourth orders for the energy density $\veps$ normalized to $T^4$ for $\Nt=6$ (left) and $\Nt=8$ (center). (Right) Entropy density for $\Nt=8$.}
\label{fig:energy}
\end{figure}

We show our results for the pressure, energy density and entropy density in Figs.~\ref{fig:pressure} and \ref{fig:energy}. Wherever possible, we have shown results for both $\Nt=6$ and 8 to emphasize that cutoff effects are small for all the observables shown here. The energy and entropy densities are obtained from the pressure from
\begin{align} &&
&& \frac{\veps}{T^4} = \sum_{n=0}^\infty \bmu{n} \left\{T\frac{dc_n}{dT} +   3  c_n\right\}   &&\text{and} 
&& \frac{s}{T^3}     = \sum_{n=0}^\infty \bmu{n} \left\{T\frac{dc_n}{dT} + (4-n)c_n\right\}.  && &&
\label{eq:thermo}
\end{align}
It is readily seen that the main correction for these values of $\muB/T$ comes from second-order susceptibilities. However fourth-order corrections do contribute, especially for temperatures in the important crossover region and lower. In the case of the energy and entropy densities (Fig.~\ref{fig:energy}), the contribution of the derivative term in Eq.~\eqref{eq:thermo} also becomes significant as the coefficient $c_4$ rises more rapidly than $c_2$ in the crossover region. The fourth-order contribution will also be larger for larger values of $\muB/T$; however for these values, it is likely that the sixth-order contribution cannot be neglected any more.


\section{Observables on the Freezeout Curve}
\label{sec:freezeout}
\begin{figure}[!htb]
\begin{center}
\includegraphics[width=0.40\textwidth]{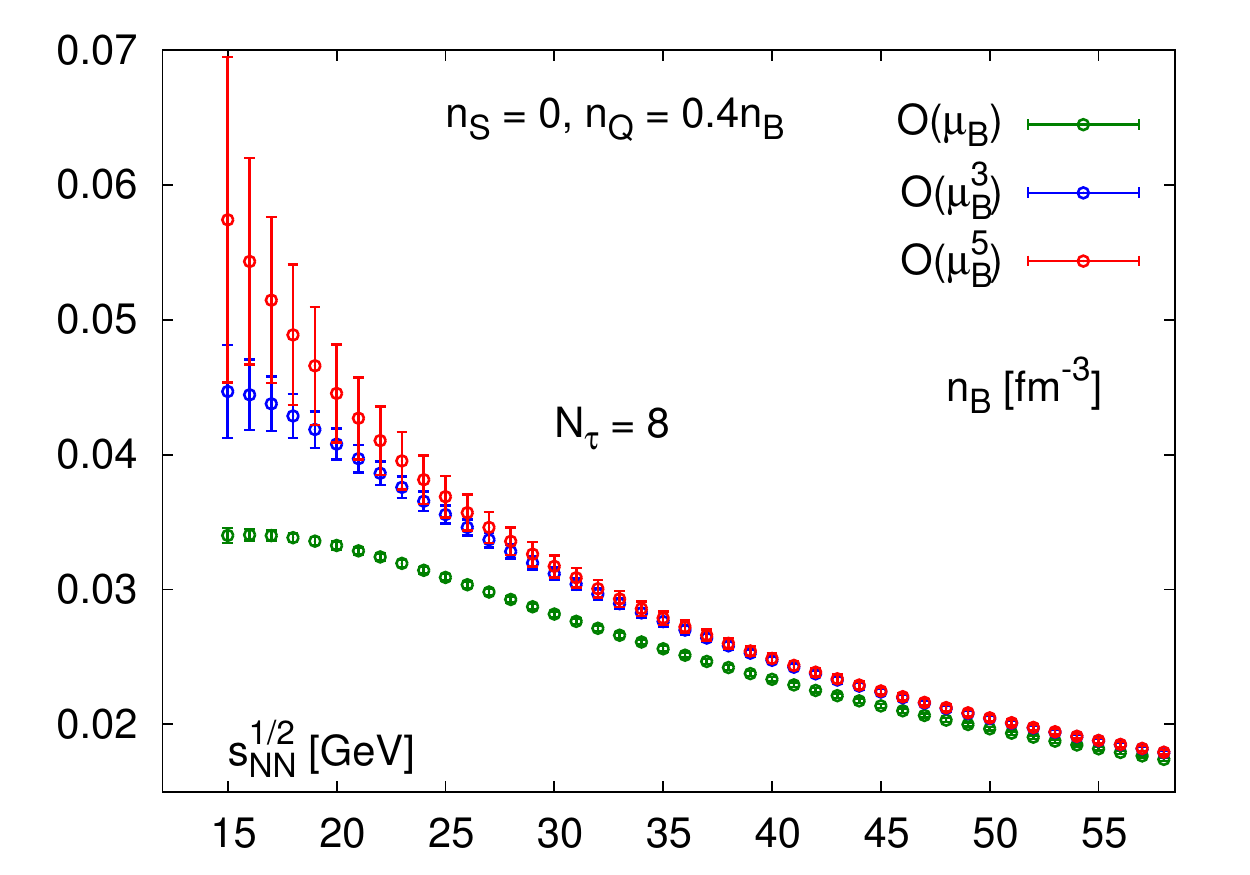}
\hspace{0.10\textwidth}%
\includegraphics[width=0.40\textwidth]{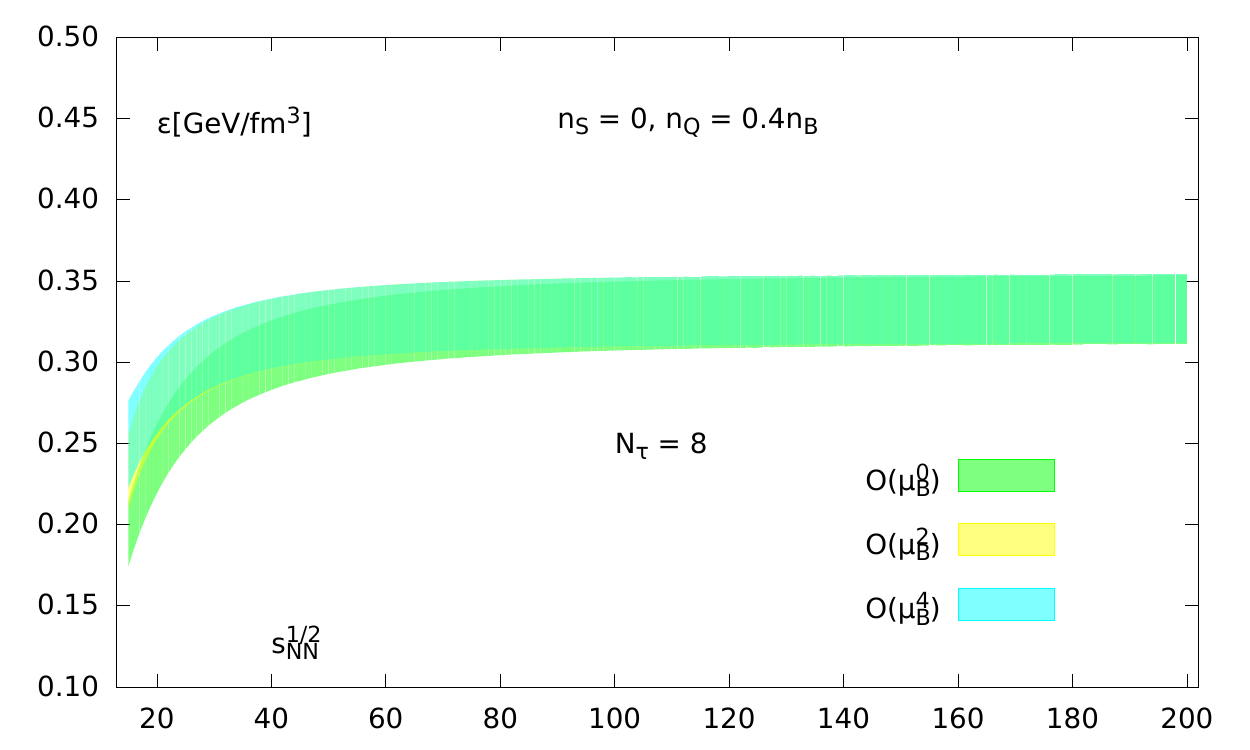}
\end{center}
\caption{(Left) The baryon density at freezeout when calculated upto leading ($\ord(\muB)$), next-to-leading ($\ord(\mu_B^3)$) and next-to-next-to-leading ($\ord(\mu_B^5)$) order. (Right) Energy density at freeze-out.}
\label{fig:freezeout}
\end{figure}

In the BES at RHIC, as the center-of-mass energy $\sNN$ is decreased, the chemical potential at freeze-out $\mu_B^f$ increases steadily while the freeze-out temperature $T^f$ changes by only about 10-15\%. While at present the EoS is known upto $\ord(\mu_B^2)$~\cite{Borsanyi:2012cr}, as one goes to lower energies, fourth-order and possibly even higher-order corrections may have to be taken into account. The determination of freeze-out parameters $T^f$ and $\mu_B^f$ is  ongoing in RHIC and LHC experiments. In particular, the estimate for the freeze-out temperature has decreased recently~\cite{Kumar:2013jxa}. In our current preliminary analysis we nonetheless use the well-known parametrization of the freeze-out curve by Cleymans \emph{et al.}~\cite{Cleymans:1999st, Karsch:2010ck} to point out some basic features of our Taylor expansion on the freeze-out curve, such as the value of $\sNN$ at which higher-order corrections start to become important.

As we saw in the previous section, in both the pressure and the energy density, the combination of zeroth and second-order terms accounted for practically the entire contribution. By contrast, the baryon number density  receives its leading contribution from $\ord(\mu_B^2)$ susceptibilities and its first corrections from the $\ord(\mu_B^4)$ ones. This makes it a good observable to study the impact of higher-order corrections.

Fig.~\ref{fig:freezeout} (left) plots the baryon number density, in units of fm$^{-3}$, on the freezeout curve as a function of the beam energy. We see that the leading-order description is a good one down to $\sNN\sim 30$ GeV, at which point the leading and next-to-leading order results start to differ. Going to still lower energies, we find that similarly, the $\ord(\mu_B^3)$ and $\ord(\mu_B^5)$ terms seem to start to differ below $\sNN\sim 15$ GeV.

A second way in which higher-order corrections might be important is seen from the energy density plot in the same figure. Freeze-out is believed to happen when the energy density drops to a certain, constant value. We can check this hypothesis by calculating the energy density on the freeze-out curve. The zeroth-order result, which only takes the temperature dependence into account, remains roughly constant down to $\sNN\sim 50$ GeV, below which it seems to dip. When $\ord(\mu_B^2)$ and $\ord(\mu_B^4)$ terms are included, the constant region is extended to slightly lower energies $\sNN\sim 30$ GeV, although the current errors preclude a more quantitative statement. Better statistics should certainly clarify the issue. However, it may be worth pointing out that the constant value of the energy density, $\veps\sim 0.3$ GeV/fm$^3$, is very close to the value in the crossover region and at the ``softest point'' of the $\muB=0$ EoS (Ref.~\cite{Bazavov:2014noa}).

\section{Conclusions}
\label{sec:conclusions}
The QCD equation of state is a necessary input in hydrodynamic models of heavy-ion collisions, and calculating it from first principles has been one of the major programs in lattice QCD. With experiments to probe nuclear matter at finite density either running (BES) or planned (BES-II and FAIR), the interest has shifted to equations of state at $\muB>0$. In this work, we presented first results from an ongoing calculation by the BNL-Bielefeld-CCNU collaboration to calculate the EoS to fourth order in the baryon chemical potential. We expect that this will eventually yield an equation of state that is valid down to beam energies of $\sNN\sim 20$ GeV and lower.

The author is partially supported by grant QLPL2014P01 of the Ministry of Education, China. The numerical calculations described here have been performed at JLab and at Indiana University in the United States and at Bielefeld University and Paderborn University in Germany. We acknowledge the support of Nvidia through the CUDA Research Center at Bielefeld University.






\end{document}